\documentclass[twocolumn,floats,a4paper,10pt]{article}

\usepackage[utf8]{inputenc}
\usepackage[USenglish]{babel}
\usepackage[]{graphicx}
\usepackage[affil-it]{authblk}
\usepackage{lipsum}
\usepackage[symbol]{footmisc}
\usepackage{hyperref}
\usepackage{amssymb}
\usepackage{amsmath}
\usepackage{cite}

\title{The dynamics of natural selection in dispersal-structured populations}

\author[1]{E. Heinsalu}
\author[1,2]{D. {Navidad Maeso}}
\author[1]{M. Patriarca\thanks{\emph{e-mail:} } }
\affil[1]{\small{National Institute of Chemical Physics and Biophysics - R{\"a}vala 10, Tallinn 15042, Estonia}}
\affil[2]{\small{Tallinn University, School of Natural Sciences and Health - Narva 29, 10120 Tallinn, Estonia}}

\begin{document}

 \twocolumn[
  \begin{@twocolumnfalse}

   \maketitle
    \begin{abstract}
    The problem of natural selection in dispersal-structured populations consisting of individuals characterized by different diffusion coefficients is studied.
The competition between the organisms is taken into account through the assumption that the reproduction and/or death probability of an individual is influenced by the number of other individuals within a neighborhood with radius $R$.
It is observed that for a wide range of parameters the competition advantage is given to the individuals whose motion is characterized by intermediate values of diffusion coefficient, instead of the most or least motile ones.
The optimal level of the dispersal is determined by the interplay between various factors such as cluster formation, temporal fluctuations, initial conditions, and carrying capacity of the system.
The dynamics and the time evolution of the system are investigated in detail revealing the winning mechanism and process of the natural selection in such dispersal-structured populations.
Furthermore, the rescaling of the results is discussed for different values of the interaction radius $R$ of the organisms. 
   \\


%PACS number(s): 87.23.Cc: Population dynamics and ecological pattern formation. \\
%      05.40.-a: Fluctuation phenomena, random processes, noise, and Brownian motion.\\
%      02.50.Ey: Stochastic processes\\

\end{abstract}
    \strut % space
  \end{@twocolumnfalse}
]

\footnotetext[1]{\emph{e-mail:} marco.patriarca@gmail.com (M. Patriarca)}

%%%%%%%%%%%%%%%%%%%%%%%%%%%%%%%%%

\section{Introduction}

%%%%%%%%%%%%%%%%%%%%%%%%%%%%%%%%%

The influence of the dispersal of  individuals on the outcome of the competition has been debated for a long time \cite{Okubo-Levin, Lewis-book}.
In some works it has been concluded that it is more advantageous to diffuse faster while in other ones the opposite conclusion has been drawn.
The picture emerging is that the temporal fluctuations, including the stochasticity induced by demographic events, tend to give a competition advantage to species diffusing faster \cite{Pigolotti-2014-PRL, Kessler-2009, Waddell-2010, Novak-2014, Johnson-1990,  Lin-2015}.
Instead, the spatial heterogeneities, due to the patch formation of organisms or non-homogeneous distribution of nutrients, give the advantage to less motile species \cite{Lin-2015, Heinsalu-2013-PRL, Hastings-1983, Holt-1985, Dockery-1998, Hutson-2003, Dieckmann-1999, Hutson-2001, Baskett-2007}, or, in the case of the species described by different types of diffusion, to the one forming stronger clusters \cite{Heinsalu-2013-PRL}.

In most investigations various explicit assumptions are made, e.g., about mutations, Allee effect, fitness, distribution of resources, carrying capacities of different space regions, costs for faster dispersal, etc.
Instead, in this paper we address a model where the spatial distribution of organisms as well as the temporal fluctuations are generated solely by the individuals themselves, and the diffusivities leading to the competition advantage are selected by this self-created environment.

In the case of the competition between two species it is straightforward to draw the conclusions: either the species diffusing faster or slower wins, or the coexistence can occur.
However, as we will demonstrate, the situation is more complex in dispersal-structured populations, in which the organisms are characterized by a wide range of diffusivities.
As discussed in Ref.~\cite{Stevens-2010}, the dispersal ability can vary as much within a species as among species, indicating that the investigation of dispersal-structured populations is highly relevant.

In accordance with the works mentioned, we observe that the general propensity is that the spatial heterogeneities tend to favor the smaller diffusivities while the increase of temporal fluctuations enhances the competition success of the individuals diffusing faster.
However, beside this general trend, we observe that in systems with moderate temporal fluctuations, instead of the utmost values, for a range of parameters the intermediate values of diffusion coefficient enhance the competition advantage, i.e., there is an optimal range of diffusion level that increases the survival probability.
The emergence of such optimal diffusivity is investigated in detail.

%%%%%%%%%%%%%%%%%%%%%%%%%%%%%%%%%

\section{Model} 

%%%%%%%%%%%%%%%%%%%%%%%%%%%%%%%%%

The model under investigation is extremely simple.
We study a system consisting of organisms that reproduce asexually, die, and move in space according to Brownian diffusion.
We assume that initially there are $N_0 = 5000$ organisms (much more than the carrying capacity of the system), placed randomly in a two-dimensional $L \times L$ square domain or one-dimensional domain with length $L$. 
The system has periodic boundary conditions and $L = 1$, so that lengths are measured in units of system size.

Differently from previous works \cite{Pigolotti-2014-PRL, Kessler-2009, Waddell-2010, Novak-2014, Johnson-1990,  Lin-2015, Heinsalu-2013-PRL, Hastings-1983, Holt-1985, Dockery-1998, Hutson-2003, Dieckmann-1999, Hutson-2001, Baskett-2007}, in the current model a heterogeneity is introduced in the population by assuming that all individuals that are present at time $t = 0$ are characterized by different diffusivities $\kappa_j$, with $j = 1, \dots, N_0$, extracted randomly from a uniform distribution in the interval $[0, 2 \kappa]$, with mean value $\kappa$ and standard deviation $\kappa /\sqrt{3}$.
Thus, the larger is $\kappa$, the larger is the variation of the individuals.
In general, it would be more realistic to assume that the diffusivities of the individuals follow the normal distribution. 
However, we have checked that using the normal distribution instead of the uniform one does not influence the results significantly, in fact, it only magnifies the effects observed.

The demographic processes are affected by the competitive interactions. 
Namely, the bug labeled $i$ [$i = 1, \dots, N$, with $N \equiv N(t)$ being the number of bugs in the system at time $t$] reproduces and dies following Poisson processes with rates $r_b^i$ and $r_d^i$ (probabilities per unit of time) \cite{EHG-2004}, respectively, 
\begin{eqnarray}
\label{rates}
\begin{aligned}
r_b^i &= \mathrm{max} (0, r_{b0} - \alpha N_R^i) \, , \\
r_d^i &= r_{d0} + \beta N_R^i \, .
\end{aligned}
\end{eqnarray}
Here, $r_{b0}$ and $r_{d0}$ are the constant reproduction and death rates of an isolated bug.
The terms containing the positive parameters $\alpha$ and $\beta$ take into account the competitive interactions: the reproduction rate of an individual $i$ decreases and the death rate increases with the number of its neighbors $N_R^i$ that are at a distance smaller than $R$ ($R \ll L$) from it.
Thus, the parameters $\alpha$, $\beta$ determine how the birth and death rates depend on the density, respectively.
The function $\mathrm{max}()$ in the first equation excludes the possibility of negative rates.
The critical number of neighbors, $N_R^*$, for which death and reproduction are equally probable for individual $i$, is determined by
\begin{equation}
\label{NR*}
N_R^* = \Delta_0/\gamma \, , 
\end{equation}
where $\Delta_0 = r_{b0} - r_{d0}$ is the maximum net growth rate and $\gamma = \alpha + \beta$ is called competition intensity.
For $N_R^i < N_R^*$ it is more probable that individual $i$ reproduces and for $N_R^i > N_R^*$ death is more likely. 
In the case of reproduction, newborns are placed at the same positions as the parents, leading to reproductive correlations, and will inherit also their characteristics (diffusion coefficient).

As mentioned, we assume that the system is initially over-crowded, but the competitive interactions will bring the population size soon down to its equilibrium size, around which  it will fluctuate.
In principle, finally --- at least after infinite time --- all organisms in the system will have the same diffusion coefficient, i.e., they are successors of the same ancestor \cite{Heinsalu-2018}.
The model described corresponds to the process of natural selection instead of evolution through mutations as investigated in numerous works, e.g., in Refs.~\cite{Novak-2014, Pigolotti-2014-PRL, Dockery-1998, Johnson-1990}.
The variety in diffusivities is assumed through the initial conditions, corresponding to the fact that the individuals are to a greater or lesser extent all different due to the natural variation and mutations. 
Thus, there is a variation in traits and our time scale is assumed to be such that mutations do not occur.
From studies of dispersal genetics it is known that the time-scale characterizing selection processes can be relevant in a wide range of size, from micro-organisms to animals and plants \cite{Saastamoinen-2018a}.

The system is simulated through the Gillespie algorithm and the spatial motion of the individuals is modeled through the continuous time random walk as described in Refs.~\cite{Heinsalu-2012-PRE, Heinsalu-2010-EPL} with the difference that now the walkers have different diffusivities and the newborns inherit the diffusivities of their parents.
%Throughout the Letter we assume that $R = 0.1$, $r_{b0} = 1$, and $r_{d0} = 0.1$.

%%%%%%%%%%%%%%%%%%%%%%%%%%%%%%%%%

\section{Identical brownian bugs} 

%%%%%%%%%%%%%%%%%%%%%%%%%%%%%%%%%

Before addressing the systems with dispersal-structured populations, let us review some aspects of the corresponding homogeneous problem, crucial for the present study.

%%%%%%%%%%%%%%%%%%%%%%%%%%%%%%%%%

\subsection{Periodic arrangement of organisms} 

%%%%%%%%%%%%%%%%%%%%%%%%%%%%%%%%%

\begin{figure}[!t]
\includegraphics[width=7.5cm]{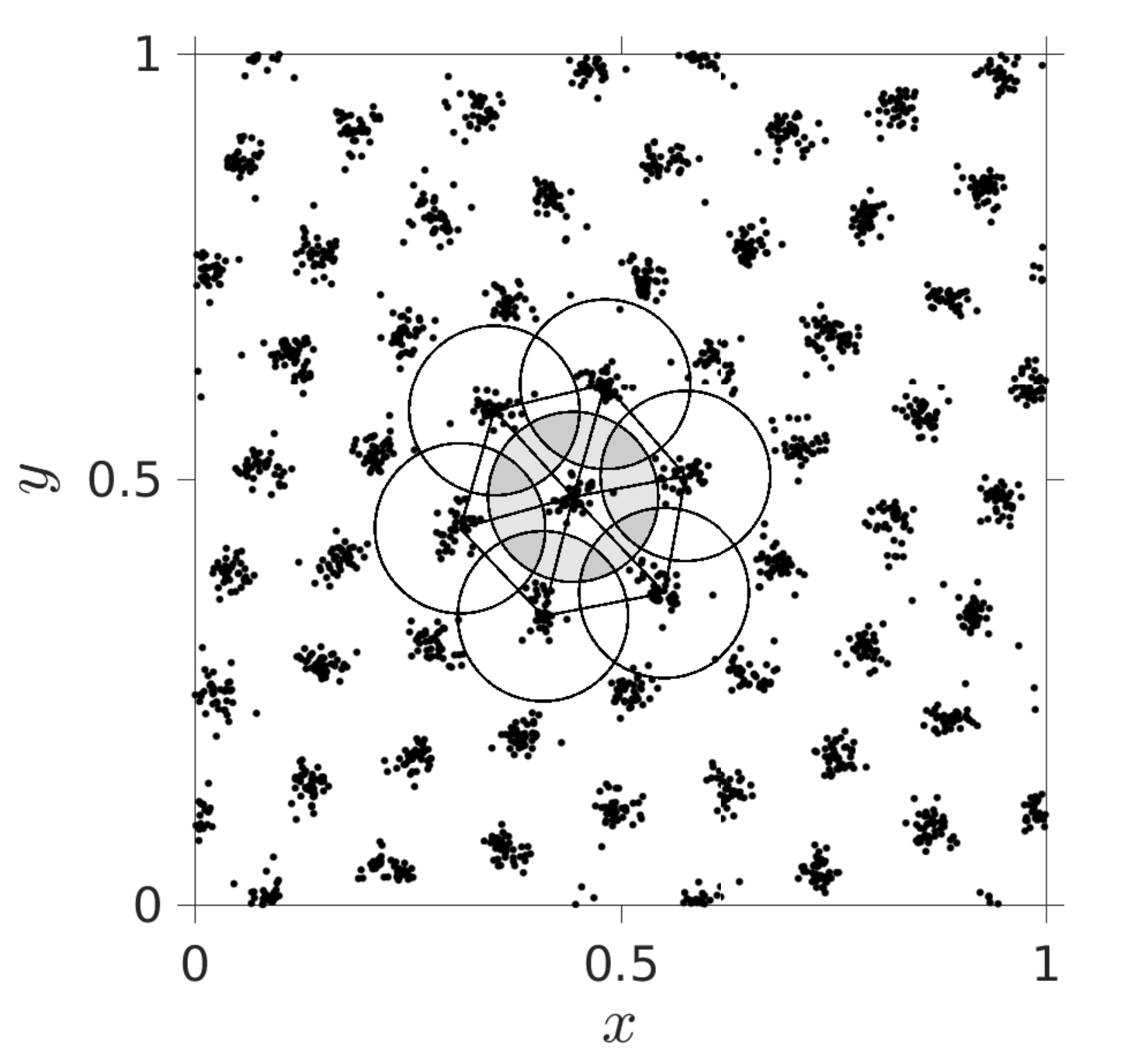}
\caption{The self-organized periodic pattern of the competing Brownian bugs with diffusion coefficient $\kappa = 10^{-5}$ in the two-dimensional homogeneous system. The circles with radius $R$ have the centers at the cluster centers. The values $R = 0.1$, $r_{b0} = 1$, $\alpha = 0.02$, $r_{d0} = 0.1$, $\beta = 0$ have been used.}
\label{XY-pattern}
\end{figure}

In the system consisting of the individuals that reproduce and die according to Eqs.~(\ref{rates}) and whose motion is characterized by the same diffusion coefficient $\kappa$, a spontaneous formation of a clumped spatial distribution of the organisms takes place under certain conditions \cite{EHG-2004, CL-2004, EHG-2015}.
In particular, in the case of a two-dimensional system with a low death rate value, a hexagonal periodic pattern appears (see Fig.~\ref{XY-pattern}).
The same takes place also when the organisms undergo L\'evy motion \cite{Heinsalu-2012-PRE, Heinsalu-2010-EPL, EHG-2015}.
The instability of the initial homogeneous spatial distribution of the organisms is a consequence of the competitive interactions.

\begin{figure}[!t]
\includegraphics[width=7.5cm]{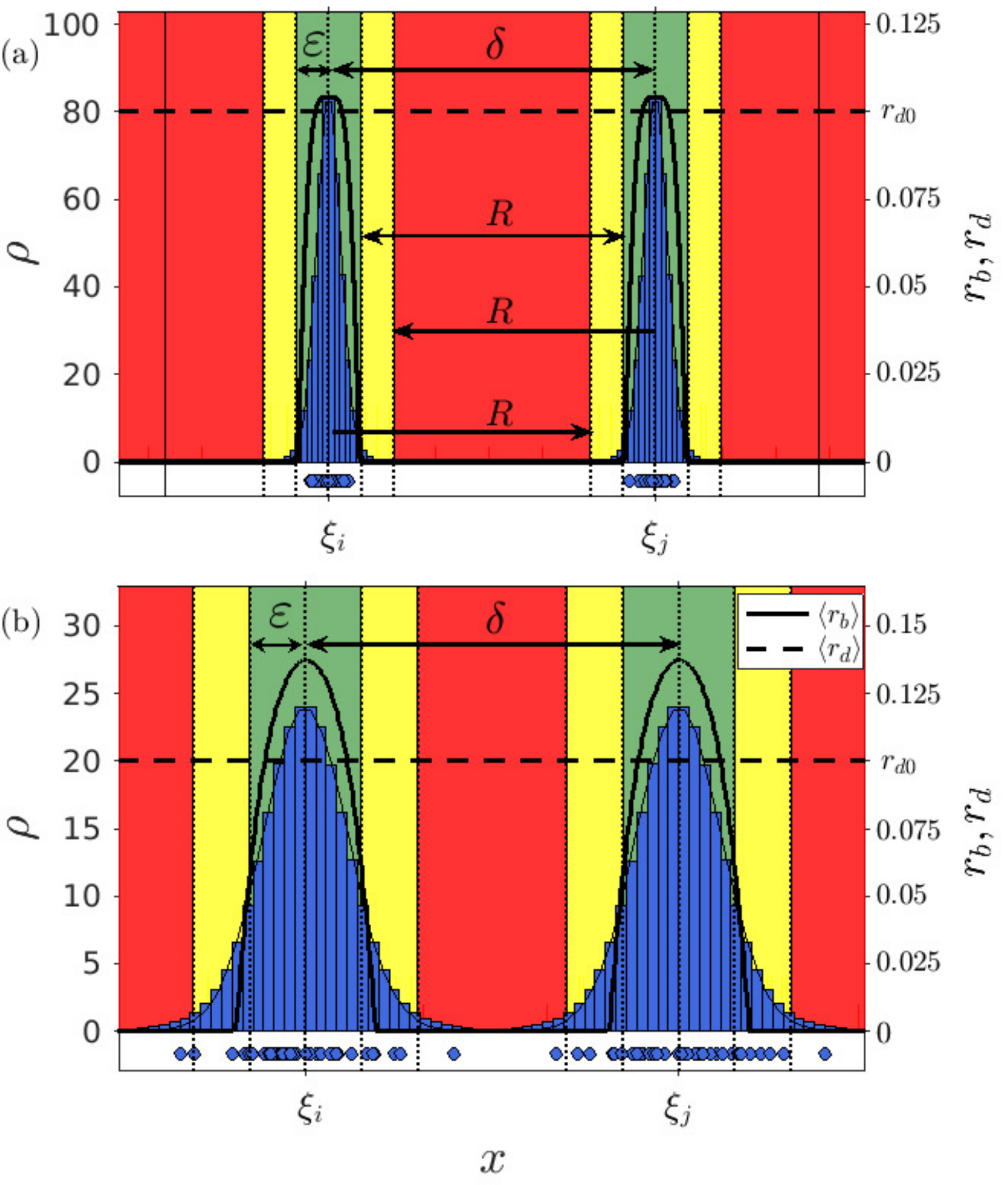}
\caption{Particle density of two neighboring average clusters in the one-dimensional homogeneous systems with (a) $\kappa = 10^{-7}$ and (b) $\kappa = 10^{-5}$; other parameter values are the same as in Fig.~\ref{XY-pattern}. The small circles below denote single Brownian bugs forming the clusters at time $t$. In average there are (a) 44 and (b) 42 bugs in a cluster. Average reproduction rate $\langle r_b \rangle$ and average death rate $\langle r_d \rangle$ depending on position $x$ inside and between two neighboring clusters has been depicted as well. Notice that the pattern periodicity $\delta$ is different for the two values of $\kappa$: (a) $\delta = 1/8$, (b) $\delta = 1/7$.}
\label{2clusters}
\end{figure}

%One can understand the instability of the homogeneous distribution by some qualitative arguments that clearly indicate that it is a consequence of the competitive interactions. 
Let the bugs be initially distributed homogeneously in space with a density such that deaths and births are balanced. 
When there are fluctuations or perturbations that enhance the density at points separated by a distance larger than $R$ but smaller than $2R$, death will become more probable than reproduction in the area in between these density maxima. 
This is so because in this zone an individual experiences the competition with the organisms from at least two of the density maxima, whereas in each density maximum the competition takes place only between the individuals in the same maximum (other maxima are out of the interaction range $R$). 
Thus, according to Eqs.~(\ref{rates}) in such region the probability for a bug to die is larger than to reproduce.
As a consequence, the density will decrease in between the density maxima. 
This in turn releases competitive pressure on the maxima, which will tend to grow, and then start to form periodically located clusters %(at a distance $fR$, with $1< f <2$) 
and close a positive feedback loop that will finally eliminate all organisms in these \textit{death zones} between the clusters (see also Fig.~\ref{2clusters}). 
%(shadowed area in Fig.~\ref{XY-pattern} for a two-dimensional system and red area in Fig.~\ref{2clusters} for a one-dimensional system). 
The only mechanism that can stop this process is diffusion, if it occurs fast enough to redistribute the bugs before the instability concentrates them \cite{EHG-2015}.

%For large values of the diffusion coefficient, the walkers appear to be distributed in an unstructured way and there is no stable pattern forming.
From the mean-field approach, using the linear stability analysis, the condition for the pattern formation is:
\begin{equation} \label{pattern}
2 R^2 \Delta_0 / \kappa > \nu_c \, , 
\end{equation}
where $\nu_c = 370.384$ for two-dimensional systems and $\nu_c = 168.4$ for one-dimensional systems. 
The derivation of Eq.~(\ref{pattern}) and of the value for $\nu_c$ is presented in great detail in Ref.~\cite{CL-2004, EHG-2015}. 

Condition~(\ref{pattern}) reveals that the pattern formation can be achieved increasing $\Delta_0$ or the interaction radius $R$ or decreasing the diffusion coefficient $\kappa$ so that $\kappa < \kappa_c$, where
\begin{equation} \label{kappaC}
\kappa_c = 2 R^2 \Delta_0 / \nu_c \, .
\end{equation}
Notice that the value of $\Delta_0$, besides satisfying Eq.~(\ref{pattern}), has to be large enough in order to avoid the situation when the system becomes extinct due to the fluctuations.

Importantly, the size of the temporal fluctuations depends not only on $\Delta_0$ and $\gamma$, determining the maximum number of organisms in the equilibrium state, but for the given values of $\Delta_0$ and $\gamma$ also on the values of $r_{d0}$ and $\beta$, as observed in Ref.~\cite{Heinsalu-2012-PRE} and discussed in Ref.~\cite{EHG-2015}.
Large temporal fluctuations lead to the situation that the clusters formed will be arranged in a rather disordered way. %; clusters may even disappear from the system.

In the following we assume that the temporal fluctuations are sufficiently small, unless indicated differently.
Throughout the paper we  set $\Delta_0 = 0.9$, $\gamma =0.02$, and $r_{d0} = 0.1$; in this case the size of the temporal fluctuations is determined by the value of $\beta$.
We also set $R = 0.1$, unless otherwise stated.

Let us first look the case of $\beta = 0$.
Then from Eq.~(\ref{pattern}) we see that for $R = 0.1$ and $\Delta_0 = 0.9$  the critical value of diffusion coefficient for pattern formation is, $\kappa_c = 4.86 \times 10^{-5}$ for two-dimensional systems and $\kappa_c = 10.69 \times 10^{-5}$ for one-dimensional systems.
The pattern periodicity (the distance between the centers of the clusters) is of the form $\delta = f R$ with $1 < f < 2$; namely, 
\begin{equation} \label{delta}
\delta \equiv 2\epsilon + R = R (2\epsilon/R + 1) = fR \, .
\end{equation}
The quantity $\epsilon$ is defined through Eq.~(\ref{delta}) and it gives an approximate estimation of the spatial cluster size, as can be seen from Fig.~\ref{2clusters} from where it is also clear that $2\epsilon < R$.
The continuous description and the linear stability analysis give that \cite{EHG-2015, EHG-2005, Cross-1993} in two-dimensional systems 
\begin{equation} \label{f-def}
f = \delta/R = 1.31475 
\end{equation}
and in one-dimensional systems 
\begin{equation} \label{f-def-1D}
f = \delta/R = 1.54 \, .  
\end{equation}

Let us illustrate the structure of the periodic pattern for the clarity in a one-dimensional system (see Fig.~\ref{2clusters}). 
According to Eq.~(\ref{NR*}) the critical number of neighbors, for which death and reproduction are equally probable for individual $i$ for the given parameters, is $N_R^* = 45$; $N_R^*+1$ is also the asymptotic equilibrium cluster size for small values of $\kappa$ ($\kappa \to 0$).
The competition in the green region in Fig.~\ref{2clusters}, i.e., around the cluster centers $\xi_{i, j}$, is smaller than in the yellow and red region; for $x \to \xi_{i,j}$ $r_b \geq r_d$.
Out of the cluster centers the organisms start to feel the competition with the ones of the neighboring clusters, $N_R^i > N_R^*$, and the probability for reproduction becomes smaller than the probability for death, $r_b < r_d$.
If the diffusion coefficient is sufficiently small then in the range $(\xi_i + \xi_j)/2 \pm R/2$ the probability for reproduction is zero (see Fig.~\ref{2clusters}); therefore, this region (the yellow and red regions together) can be called death zone.
In fact, defining the death zone through the condition $r_d > r_b$ then, as can be seen from Fig.~\ref{2clusters}, it is even wider than $R$.
Notice that in earlier works the region $(\xi_j - R, \xi_i + R)$ has been called the death zone (the red zone in Fig.~\ref{2clusters}, $\xi_i$, $\xi_j$ are the centers of two neighboring clusters); in this region the competition is extreme because the individuals feel very high competition compared to the ones inside the cluster and the density of organisms is zero. 

In the two-dimensional system, besides the zones where the organisms feel the competition pressure of two neighboring clusters,  there are also zones where the competition from three clusters is felt.
However, to pass from one cluster to the next one, there are narrow channels  where the influence of only two neighboring clusters is felt.
Such super-competition regions and lower competition channels (see Fig.~\ref{XY-pattern}) allow to reduce the problem of diffusing from one cluster to the next one to the quasi one-dimensional problem.

As already mentioned, for $\kappa \to 0$ the number of organisms inside clusters reaches the value $N_R^*+1$.
In this case the probability for reproduction and death is equal in the cluster centers.
For $\kappa > 0$ in the cluster centers the probability for reproduction is  slightly larger than the probability for death and the net growth rate $r_b - r_d$ in the cluster centers increases when the diffusion coefficient increases due to the decrease of the number of organisms inside the clusters and even more due to the spreading of clusters (compare panels a and b in Fig.~\ref{2clusters}). 
This is similar to the idea of kin competition \cite{Hamilton-1977, Comins-1980, Frank-1986, Gandon-1999, Heino-2001, Dyken-2010} --- leaving the birth place (clusters) relieves the competition inside the clusters, creating a more favorable environment inside the clusters compared to the zero diffusion case when everybody would remain there; for individuals leaving the clusters the situation is of course not advantageous within the current model.
Furthermore, the total number of organisms in the system is decreased compared to the low diffusion case.
For large values of diffusion the clusters disappear and one cannot talk about the kin competition anymore.

%%%%%%%%%%%%%%%%%%%%%%%%%%%%%%%%%

\subsubsection{Invading neighboring clusters} 

%%%%%%%%%%%%%%%%%%%%%%%%%%%%%%%%%

%It is known that the kin competition is also a major driving force for invasions.

In Ref.~\cite{Heinsalu-2012-PRE} the bugs with the same diffusion coefficient were divided into different groups according to their initial position and the mixing of these groups was investigated.
It was observed that in the case of small diffusivities the mixing of groups did not take place or if then only due to the diffusion of clusters as a whole during the clusters arrangement into the periodic pattern. 
For larger values of $\kappa$ the inter-cluster travel took place and led to the conquering of new territories; i.e., bugs were found in a region where their ancestors were not from. 
The effect was larger for larger $\kappa$ and led to the disappearance of some initially present groups. 
Finally, for increased diffusion, due to the intra-cluster competition all surviving bugs were from a single group (and finally from a single ancestor); which group (ancestor) won was a random event. 
The process was faster for larger diffusion.

The diffusion coefficient that makes it possible to traverse the death-zones between two neighboring clusters can be estimated from the following condition:
\begin{equation} \label{max}
t^{*} = t_m \, .
\end{equation}
Here $t^*$ is the typical first-passage time of an organism with the diffusion coefficient $\kappa$ for traversing the death-zone between the clusters. %diffusing from the center of one cluster to the center of the other cluster.
And $t_m$ is the typical lifetime of a family defined as a chosen individual and its descendants.
Thus, even if an individual with diffusivity $\kappa$ does not reach a neighboring cluster, its descendants that continue the diffusion process of the mother may arrive there and in this way the organisms with a certain $\kappa$ can invade new clusters even if $\kappa$ is rather small (so that the probability for a single organism to arrive there is extremely small).

For simplicity, let us consider a one-dimensional system and assume that $\beta = 0$ (for $\beta > 0$ the calculation of the family life time is not trivial). 

The width of the death-zone is approximately the interaction radius $R$ (the yellow and red region in Fig.~\ref{2clusters}) and the typical first-passage time of an organism with the diffusion coefficient $\kappa$ for traversing the death-zone between the clusters is $t^* = R^2/(6 \kappa)$.

However, in order to invade the neighboring cluster traversing the death-zone might not be sufficient. 
%i.e., in the case of sufficiently small diffusion coefficient.
As can be seen from Fig.~\ref{2clusters}, unless the diffusion coefficient is really small, the bugs extend due to the diffusion also to the death-zone. 
Therefore, the individuals on the border may still experience the competition from more than just one cluster: their reproduction rate is decreased and the death rate is larger than the birth rate, though they are inside the cluster (in the green region).
Furthermore, one has to be successful also in the intra-cluster competition. 
%When the clusters extend also into the yellow part of the death-zones due to the diffusion, the bugs that are near or on the border of the death zones feel also the influence of the neighboring cluster.
%In this case, the distance that the bugs have to travel in order to arrive from the safe zone to the safe zone  where the probability for reproduction is higher than the probability for the death, is larger than $R$ and can be estimated approximately equal to $\delta$.
The probability for reproduction is maximal in the cluster centers and the difference $r_b - r_d$ is the larger the larger is the diffusion coefficient.
Then, the distance $\ell$ that the individuals have to traverse for successful invasion can be %estimated approximately 
equal even to $\delta$.

Thus, in Eq.~(\ref{max}) the following typical first-passage time should be used,
\begin{equation}
\label{t*}
t^* = \ell^2/(6 \kappa)  \, ,
\end{equation}
where $\ell \in [R, \delta]$.

Following Ref.~\cite{EHG-2005} one finds that the  the typical lifetime of a family is,
\begin{equation}
\label{tm}
t_m = \Delta_0/(2 \alpha r_{d0})  \, .
\end{equation}

Thus, from conditions~(\ref{max}), (\ref{t*}), (\ref{tm}) we get that the critical diffusion coefficient allowing to traverse the inter-cluster death-zones is
\begin{equation}
\label{kappaopt}
{\kappa}^* = \ell^2 \alpha r_{d0} / (3 \Delta_0)  \, .
\end{equation}
%
%For the parameter values used in Figs.~\ref{XY-pattern} and \ref{2clusters}  
%%Assuming that $R = 0.1$, $\Delta_0 = 0.9$, $\alpha = 0.02$, $r_{d0} = 0.1$, $\beta = 0$,  
%we get that ${\kappa}^* = 0.74 \times 10^{-5}$.

%Considering Eq.~(\ref{f-def-1D}), for our parameters we have then that the critical diffusion coefficient leading to the successful invasion is ${\kappa}^* = 1.76 \times 10^{-5}$.

The one-dimensional approximation is sufficiently good also for a two-dimensional system, as discussed above, but then one should use Eq.~(\ref{f-def}); as a result ${\kappa}^* = 1.28 \times 10^{-5}$ in two-dimensional systems.
The parameter values used for Fig.~8 in Ref.~\cite{Heinsalu-2012-PRE}  %($R = 0.1$, $\Delta_0 = 0.9$, $\alpha = 0.02$, $r_{d0} = 0.1$, $\beta = 0$) we have that 
are the same as we have used here.
Thus, also in this case it should be true that ${\kappa_j}^* \approx 1.28 \times 10^{-5}$. 
This result is in consistency with what is observed in Fig.~8 in Ref.~\cite{Heinsalu-2012-PRE}: for $\kappa = 10^{-5}$ the mixing of different groups is still not visible, but for $\kappa = 2 \times 10^{-5} > \kappa^{*}$ it is already rather noticeable.

We also point out that Eqs.~(\ref{f-def}) and (\ref{f-def-1D}) for the pattern periodicity hold only close enough to the instability developing in a periodic pattern.
Numerical simulations reveal that for fixed parameter values the diffusion coefficient influences to a certain extent the periodicity of the pattern (see Fig.~\ref{2clusters}) and the number of clusters in the system: smaller values of $\kappa$ lead to larger number of clusters. 
For example, from Fig.~8 in Ref.~\cite{Heinsalu-2012-PRE} one can see that the number of clusters is $53$ for $\kappa = 4 \times 10^{-5}$ and $63$ for $\kappa = 10^{-5}$.
A one-dimensional system with $L = 1$ can fit $8$ clusters for $\kappa = 10^{-6}$, and thus $\delta = 0.125$; for $\kappa = 10^{-5}$ and for $\kappa = 10^{-4}$ the system fits $7$ or $8$ clusters leading to $\delta \approx 0.143$ or $\delta = 0.125$; the actual distances between the cluster centers fluctuate around these values.
The dependence of the pattern periodicity on diffusion coefficient enters through the cluster linear size dependence on the diffusion coefficient, but the mean-field description fails in describing this.
Furthermore, also the size of the simulation domain affects the inter-cluster distance; this issue will be addressed in detail in a forthcoming paper.

\section{Natural selection in dispersal-structured populations}

%%%%%%%%%%%%%%%%%%%%%%%%%%%%%%%%%%%%%%%%%%%%%

Let us now go back to the problem of dispersal-structured populations.

In the following, as already mentioned, we assume that $r_{b0} = 1$, $r_{d0} = 0.1$, and $\gamma = 0.02$; $R = 0.1$, unless indicated differently.

First, we investigate the system where the death rate is constant and only the birth rate is influenced by the competition, e.g., we set $\alpha = 0.02$ and $\beta = 0$.
For such parameter values the temporal fluctuations are rather small.
Then we investigate how the increase of temporal fluctuations influences the process of natural selection in the dispersal-structured populations.
Finally, we also study the effect of the interaction radius size on the competition outcome.

%%%%%%%%%%%%%%%%%%%%%%%%%%%%%%%%%%%%%%%%%%%%%

\subsection{Small temporal fluctuations} 

%%%%%%%%%%%%%%%%%%%%%%%%%%%%%%%%%%%%%%%%%%%%%

%%%%%%%%%%%%%%%%%%%%%%%%%%%%%%%%%%%%%%%%%%%%%

\subsubsection{The dynamics of competition: time evolution}  \label{Sec-evolution}

%%%%%%%%%%%%%%%%%%%%%%%%%%%%%%%%%%%%%%%%%%%%%

%
\begin{figure}[!t]
\includegraphics[width=7.5cm]{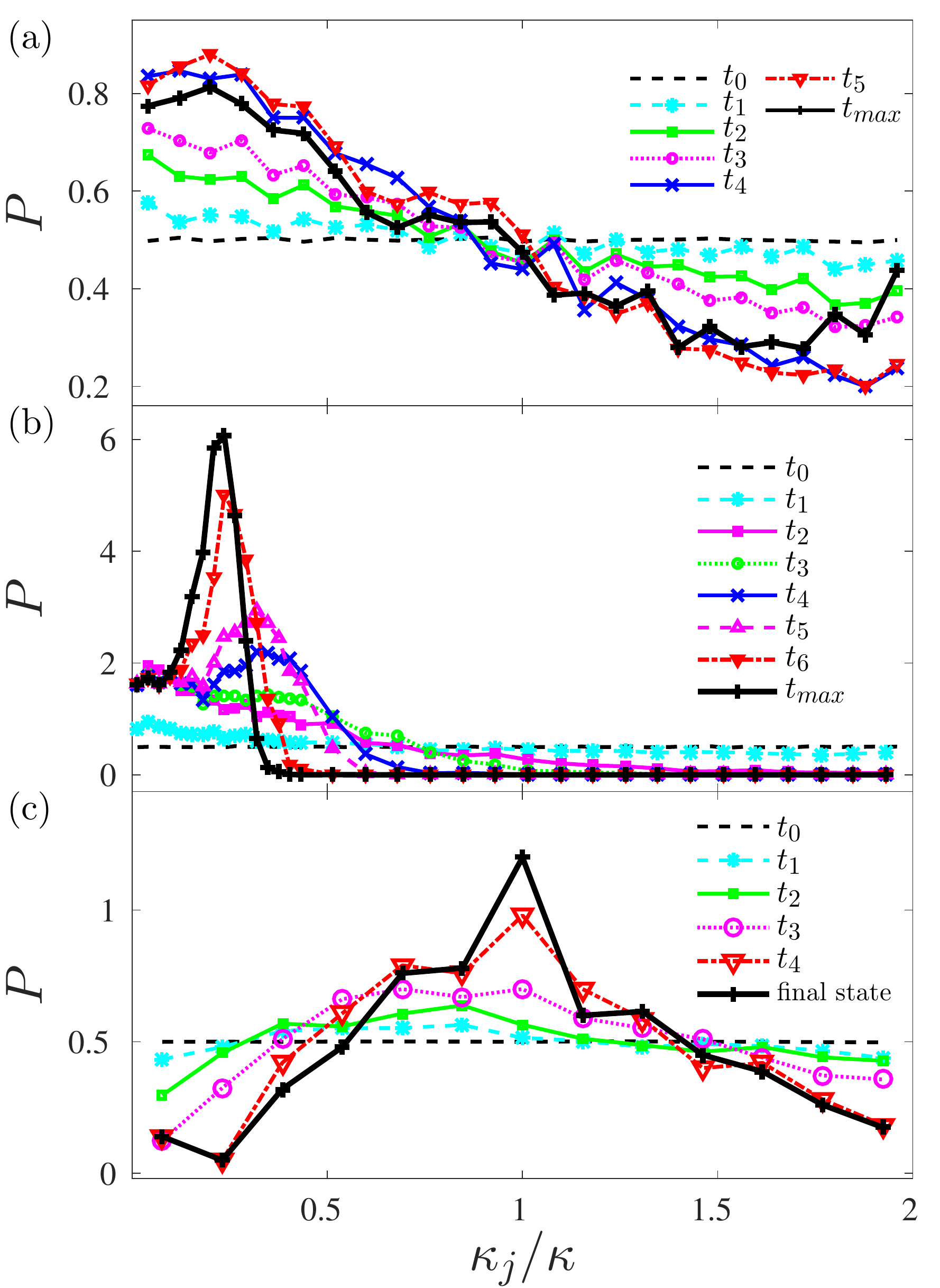}
\caption{The time evolution of the probability distribution $P(\kappa_j)$ in the heterogeneous two-dimensional system ($\beta = 0$, $\alpha = 0.02$). The initial mean diffusion coefficients are: (a) $\kappa = 10^{-5}$; (b) $\kappa = 10^{-4}$; (c) $\kappa = 10^{-3}$. Notice that the x-axes is scaled by different values of $\kappa$ for the three panels. The curves are obtained averaging over $150$ realizations.}
\label{tevol-b0}
\end{figure}

In the case of the dispersal-structured populations, there are initially $N_0$ organisms all with different diffusivities (remember that we assumed that $N_0$ is much larger than the carrying capacity of the system). 
Due to the fluctuations in the number of individuals and the irreversibility of death, the number of different diffusivities decreases in time, reaching, in principle, after a certain time for any parameters the value $1$.
The required time is the larger the smaller is the mean value $\kappa$ of the initial diffusivities and we have observed previously that for some parameter values the disappearance times of the diffusion coefficients diverge \cite{Heinsalu-2018, Heinsalu-2013-PRL}.
However, considering that in real systems the time is always finite and because we actually investigate a process taking place in a limited time interval --- smaller than the mutation time scale --- we have set a maximum simulation time $t_\mathrm{max} = 5 \times 10^{5}$.
Thus, the probability distribution $P(\kappa_j)$ in the final state is constructed either on the basis of the global diffusion coefficients, i.e., in single realizations all individuals have finally the same diffusivities, meaning that they are all the successors of the same ancestor, or on the basis of the different diffusivities present in the system at $t_\mathrm{max}$.

In the finite time interval the two-dimensional system is most selective in the diffusivities for intermediate values of $\kappa$ (e.g., for $\kappa = 10^{-4}$).
In this case, $P(\kappa_j)$ presents at $t_\mathrm{max}$ a very clear maximum at smaller but intermediate values, going then rather rapidly to zero; the distribution has a finite value  at $\kappa_j \to 0$ (see the curve corresponding to $t_\mathrm{max}$ in Fig.~\ref{tevol-b0}b).
The time evolution in Fig.~\ref{tevol-b0}b shows that in the beginning the organisms with the smallest diffusivities are the most favored ones due to the faster density enhancement near the individuals with smaller $\kappa_j$.
Related with the inter-cluster competition the second maximum in $P(\kappa_j)$ appears at $\kappa_j$ values close to the critical value determined by Eq.~(\ref{pattern}) leading to the pattern formation. %; for the given parameters the critical diffusion coefficient leading to the instability developing in a periodic pattern is $\kappa_c \approx 4.86 \times 10^{-5}$.
What happens is that the organisms with $\kappa_j \approx \kappa_c$, still forming into clear clumps in the case of an unstructured population, invade the space regions occupied by the organisms with a comparable but slightly larger diffusion coefficients.
As this process goes on in time, the distribution $P(\kappa_j)$ becomes narrower, the initially local maximum of the distribution increases, turns after some time into the global maximum, and shifts to the smaller values while the individuals with larger diffusivities disappear gradually.
At the same time, the value of $P(\kappa_j \to 0)$ decreases gradually: the individuals with $\kappa_j \to 0$ have an advantage in the intra-cluster competition, but the probability that they manage to traverse the zones between the clusters, where the probability for the death is larger than for the reproduction \cite{EHG-2015, Heinsalu-2012-PRE}, is very low; the irreversibility of death leads finally to disappearance of low diffusivities (not apparent in Fig.~\ref{tevol-b0}b).
Thus, the maximum of $P(\kappa_j)$ at intermediate values is the outcome of the inter- and intra-cluster competition.

For small values of $\kappa$, e.g., for $\kappa = 10^{-5}$, corresponding to the situation when each value of $\kappa_j$ leads to the strong clustering, at $t_\mathrm{max}$ (almost) each patch of the periodic pattern emerging is occupied by individuals coming from a different ancestor (see also Ref.~\cite{Heinsalu-2018}).
The behavior of $P(\kappa_j)$ is similar to the case of intermediate values of $\kappa$: smaller diffusivities tend to favor the competition success, but the maximum of the probability distribution of $\kappa_j$ is at an intermediate value of $\kappa_j$ (see Fig.~\ref{tevol-b0}a). 
However, now $P(\kappa_j)$ has a finite value also for $\kappa_j \to 2 \kappa$, i.e., also the species with larger diffusivities manage to survive at large but finite times.
The reason why in the case of small $\kappa$ the initial and final distributions of $\kappa_j$ are rather similar within the finite time, is that the variation in diffusivities is small and the probability to traverse the death-zones is very low for all values of $\kappa_j$; this makes the whole dynamics very slow.

For large values of $\kappa$, e.g., for $\kappa = 10^{-3}$, the time evolution of $P(\kappa_j)$ changes significantly (see Fig.~\ref{tevol-b0}c).
In this case, the small values of $\kappa_j$ never become the favorable ones during the time evolution.
Instead, starting with an uniform distribution, as time passes, the probability distribution $P(\kappa_j)$ will decrease at $\kappa_j \to 0$ as well as at $\kappa_j \to 2 \kappa$, i.e., both smallest as well as largest diffusivities are disadvantageous.
Simultaneously, $P(\kappa_j)$ develops a maximum at intermediate values of $\kappa_j$.
The disadvantage of the large diffusivities is related with the spatial inhomogeneities (still present for the given $\kappa$) caused by the reproductive correlations. 
The disadvantage of the small diffusivities is caused by the initial condition leading to a small fraction of individuals with small $\kappa_j$ and a large fraction of individuals with larger $\kappa_j$ that create a well-mixed environment and prevent the successful cluster formation of the organisms with small $\kappa_j$ who should have the competition advantage. 
We have checked that assuming an initial distribution where the small values of $\kappa_j$ have a sufficiently larger probability, e.g., a truncated exponential distribution in the same interval $[0, 2 \kappa]$ (notice that the mean value is now different), also the final distribution has a maximum at small values of $\kappa_j$.

%%%%%%%%%%%%%%%%%%%%%%%%%%%%%%%%%%%%%%%%%%%%%

\subsubsection{One-dimensional systems} 

%%%%%%%%%%%%%%%%%%%%%%%%%%%%%%%%%%%%%%%%%%%%%

%
\begin{figure}[!t]
\includegraphics[width=7.5cm]{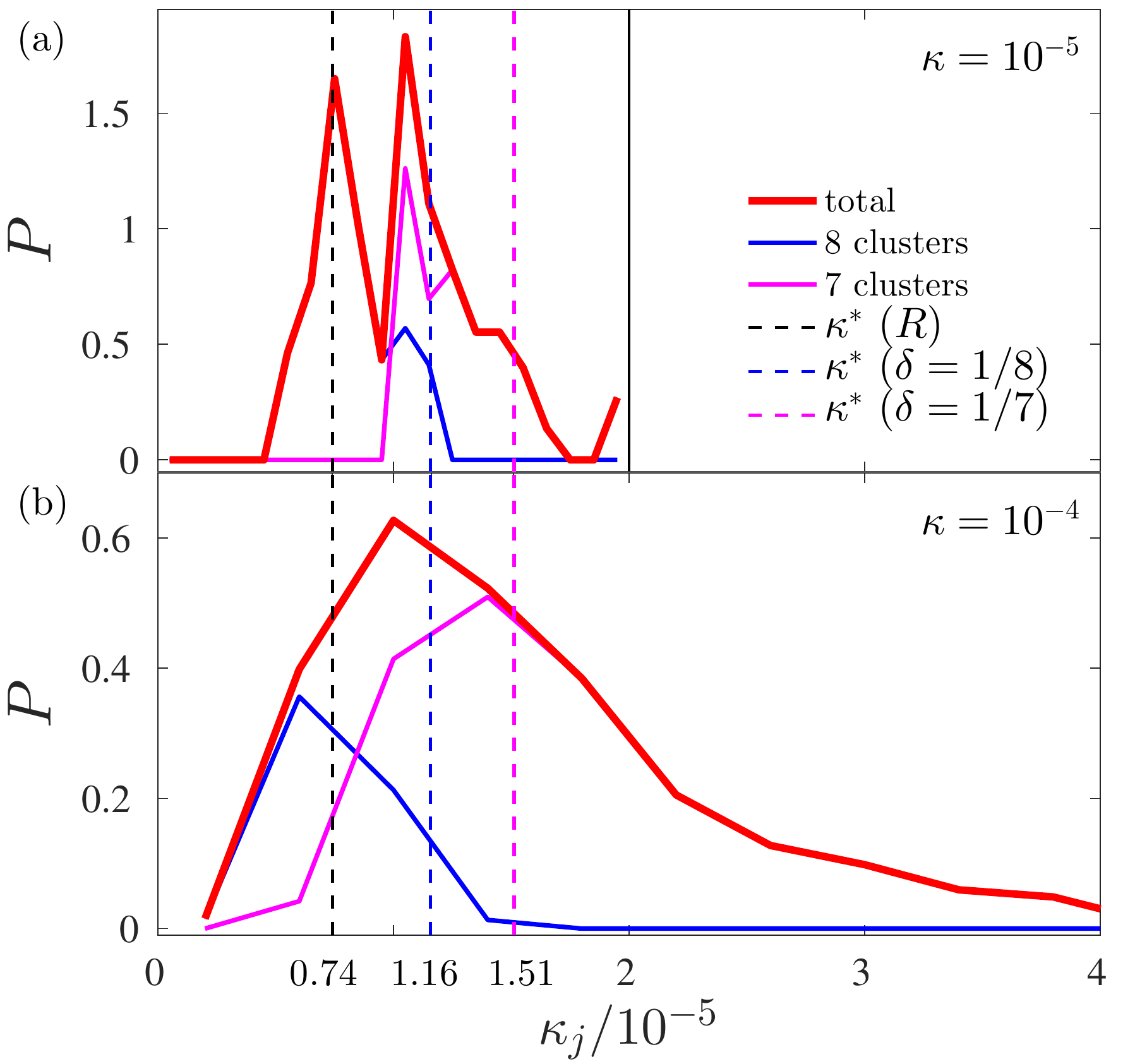}
\caption{The final probability distributions $P(\kappa_j)$  in the heterogeneous one-dimensional system ($\beta = 0$, $\alpha = 0.02$).  
The initial mean diffusion coefficients are: (a) $\kappa = 10^{-5}$; (b) $\kappa = 10^{-4}$.
The solid red curves take into account all the realizations.
Instead, the blue and pink dashed curves are obtained when separating the realizations leading to 8 or 7 clusters, respectively.
The vertical dashed lines from left to right represent the theoretical results for $\kappa^{*}$ from Eq.~(\ref{kappaopt}) for $\ell = R$ (the first curve) and $\ell = \delta$, taking into account that there are $8$ and $7$ clusters in the system, i.e., $\delta = 0.125$ and $\delta = 0.143$.}
\label{histogram-1dim-4}
\end{figure}

As mentioned, in the case of small and intermediate values of $\kappa$, due to the very large disappearance times \cite{Heinsalu-2018}, we cannot reach the state of the system when finally there is only one $\kappa_j$ present.
Thus, one can question what is the final probability distribution of $\kappa_j$, because it might still change significantly compared to the ones in Figs.~\ref{tevol-b0}a and \ref{tevol-b0}b at $t_\mathrm{max}$.
From Figs.~\ref{tevol-b0}a and \ref{tevol-b0}b it seems that a stabilizing selection takes place and that there is an optimal diffusivity range leading to the increase of the competition success.
However, this is not what is predicted by the mean-field theory.
According to the latter one, the directional selection should take place \cite{Heinsalu-2013-PRL,EHG-2015,Hastings-1983,Dockery-1998}, i.e. in the case of small temporal fluctuations the smallest diffusivities win while large temporal fluctuations give the advantage to the organisms diffusing faster.
%However, based on the results obtained for the three ranges of $\kappa$, we deduce that, if $\beta = 0$ so that the temporal fluctuations in the system are small, as a consequence of the interplay between the inter- and intra-cluster competition, also the final probability distribution of $\kappa_j$ has a maximum at an intermediate value $0 < \kappa_j < 2 \kappa$. 

In order to find the answer whether a stabilizing selection takes place or not, we have investigated the one-dimensional systems that converge to the final state when all the organisms have the same diffusion coefficient within an accessible simulation time.

For the dispersal-structured system with the initial mean diffusion coefficient $\kappa = 10^{-4}$ we observe in Fig.~\ref{histogram-1dim-4}b (solid red curve) that the probability distribution of the diffusivities $\kappa_j$ in the final state shows a clear maximum at intermediate values.
For the initial mean diffusion coefficient $\kappa = 10^{-5}$ the probability distribution of the diffusivities $\kappa_j$ in the final state demonstrates, instead, two maxima at intermediate values, see Fig.~\ref{histogram-1dim-4}a (solid red curve).
In both cases the enhancement of $P(\kappa_j)$ is approximately in the same range of $\kappa_j$ (notice that differently from Fig.~\ref{tevol-b0} in Fig.~\ref{histogram-1dim-4} both panels are rescaled by the same value of $\kappa$).
Thus, there really exists a range of optimal diffusivities giving the competition advantage, i.e. stabilizing selection, and it is determined by $\kappa^*$, as will be discussed in the following.

As mentioned above, analyzing the systems we have found that in both cases, for $\kappa = 10^{-5}$ as well as for $\kappa = 10^{-4}$, the organisms can self-organize in 7 or 8 clusters. 
Decomposing the probability distribution $P(\kappa_j)$ correspondingly, we see from Fig.~\ref{histogram-1dim-4}a that in the case of $\kappa = 10^{-5}$ the first maximum of the total distribution is related to the systems where 8 clusters are formed and the second maximum is related to the systems where 7 clusters are formed.
The probability to have in the final state 7 or 8 clusters is $50/50$; however, in the beginning of the time evolution the probability to have 7 clusters is much higher (around $80\%$).
In the case of $\kappa = 10^{-4}$ the probability to have 8 clusters is much lower (less than $5\%$ at small times and around $20\%$ in the final state) and the maxima of the two sub-distributions cannot be resolved, i.e. the total distribution has a single maximum (see Fig.~\ref{histogram-1dim-4}b).

Though in the case of 8 clusters in average $\delta = 0.125$, the probability distributions have the maximum around the value $\kappa^* = 0.74 \times 10^{-5}$ corresponding to the distance $\ell = R=0.1$ (Eq.~(\ref{kappaopt})), i.e. it is sufficient to arrive from a cluster border to a cluster border.
The probability distributions for the systems where 7 clusters are formed (i.e. when in average $\delta = 0.143$) have a maximum around the value determined by Eq.~(\ref{kappaopt}) with $\ell = \delta$; however, notice that while for $\kappa = 10^{-4}$ the maximum is approximately around the critical diffusion coefficient $\kappa^* = 1.51 \times 10^{-5}$ corresponding to $\delta = 0.143$ (7 clusters), for $\kappa = 10^{-5}$ the maximum is shifted towards smaller values; for $\kappa = 10^{-4}$ the distribution for the systems with 7 clusters is also remarkably broader.

As proposed in Sec.~\ref{Sec-evolution}, the optimal diffusivity is determined by the interplay between the inter- and intra-cluster competition.
The slight difference in the position of the maximum for the probability distribution $P(\kappa_j)$ for the systems with 7 clusters in the case of $\kappa = 10^{-5}$ and $\kappa = 10^{-4}$ is related to the intra-cluster competition.
Because in the case of larger diffusivities in the cluster centers the reproduction rate is much higher, as can be seen from Fig.~\ref{2clusters}, then the distance to be traversed is from cluster center to cluster center, as discussed above, i.e. larger diffusion coefficient is needed to be successful in both the inter- as well as in the intra-cluster competition.
In the case of smaller diffusivities it is not needed to travel from cluster center to cluster center but a smaller distance ($ R < l < \delta$) is sufficient.

%According to Sec.~\ref{Sec-invasion} the critical diffusion coefficient enabling a family to traverse the death-zone is given by Eq.~(\ref{kappaopt-delta}).
%Fig.~\ref{histogram-1dim-4} demonstrates that the theoretical prediction made in Sec.~\ref{Sec-invasion} agrees well with the numerical result.
%The organisms whose motion is characterized approximately with such value of diffusion coefficient have the intra-cluster competition advantage respect to the organisms diffusing considerably faster and inter-cluster competition advantage respect to the organisms diffusing considerably slower so that they are not capable to invade the neighboring clusters.

Thus, on the basis of the one-dimensional simulations one can conclude that there exists an optimal range of diffusivities giving the competition advantage.
It is determined by the inter- and intra-cluster competition, and therefore by $\kappa^*$, and depends also on whether the organisms self-organize into 7 or 8 clusters in the system and slightly on the initial mean diffusion coefficient $\kappa$ of the system, as well as on fluctuations determined for $\beta =0$ by $r_{d0}$: the smaller is $r_{d0}$, the smaller are fluctuations and in agreement with Eq.~(\ref{kappaopt}) the maximum of $P(\kappa_j)$ shifts to smaller values of $\kappa_j$ in consistency with the mean-field approximation. %\cite{Heinsalu-2013-PRL, Hastings-1983, Dockery-1998}.
 
The situation would be different in a system where the individuals perform L\'evy walks.
In this case, due to the occasional long jumps it would be always possible to arrive to the other clusters and the outcome would be determined solely by the intra-cluster competition \cite{Heinsalu-2012-PRE}.
Thus, in this case there would take place directional selection and the slowest diffusing organisms would have the competition advantage.
The same is valid when taking into account the mutation process.
%In these cases it would be always the species with the smallest diffusion coefficient to win the competition.

%ADD HERE THE FIGURES FOR KAPPA$=10^{-5}$ FOR DIFFERENT RANGES - 1D HOMOGENEOUS CASE!!! ???

%%%%%%%%%%%%%%%%%%%%%%%%%%%%%%%%%%%%%%%%%%%%%

\subsection{The influence of increasing temporal fluctuations} 

%%%%%%%%%%%%%%%%%%%%%%%%%%%%%%%%%%%%%%%%%%%%%

We now return to the two-dimensional case and investigate the system where also the death rates are neighborhood dependent, i.e., $\beta > 0$. 
In order to have a comparable situation respect to the case $\beta = 0$, we keep the sum $\alpha + \beta = \mathrm{const.}$, i.e., the critical number of neighbors $N_R^*$ of particle $i$ is always the same, see Eq.~(\ref{NR*}).
If the individuals are identical, the increase of $\beta$ leads to the increase of the fluctuations amplitude in the population size.
The spatial distribution becomes for a given diffusion coefficient more clumped. 
The regular hexagonal pattern appearing for $\beta = 0$ at sufficiently low diffusion coefficients gets for $\beta > 0$ irregular and the centers of mass of the clusters will rather perform a random walk instead of fluctuating slightly around the fixed positions of the pattern, and occasionally disappear from the system \cite{EHG-2015}.

\begin{figure}[!t]
\includegraphics[width=7.5cm]{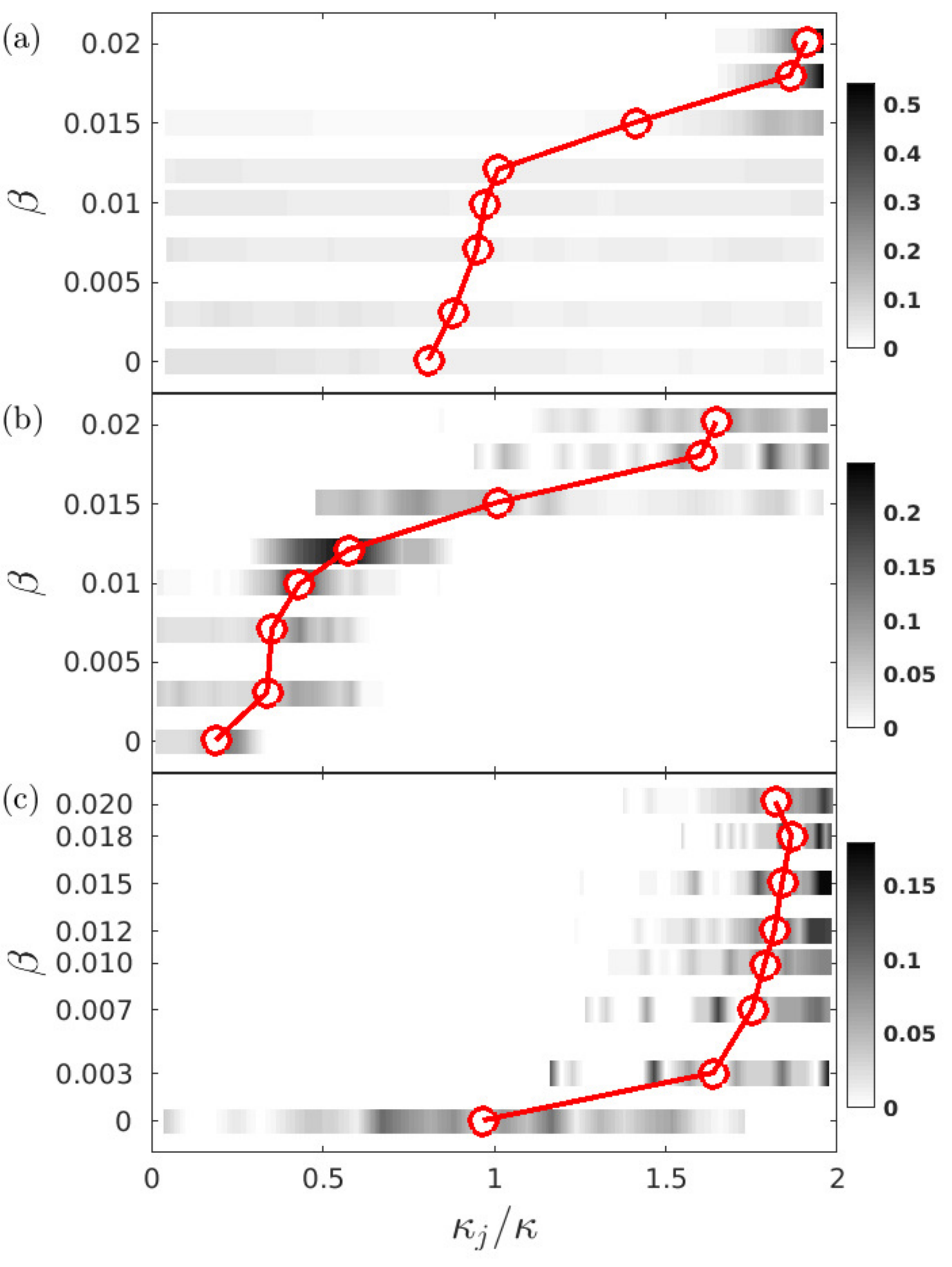}
\caption{Diffusion coefficients $\kappa_j$ leading to the competition success in the two-dimensional system for different values of $\beta$ ($\alpha = 0.02 - \beta$) --- the projection of the probability distribution $P(\kappa_j)$ in the final state (or at $t_\mathrm{max}$); the darker the color the higher is the probability, as indicated by the legend. The initial mean diffusion coefficients are: (a) $\kappa = 10^{-5}$, (b) $\kappa = 10^{-4}$, (c) $\kappa = 10^{-3}$. The curves are obtained averaging over $150$ realizations. The circles indicate the mean diffusion coefficients of the systems in the final states (or at $t_\mathrm{max}$).}
\label{beta}
\end{figure}

In the case of the dispersal-structured populations the results concerning the competition between different diffusivities are illustrated by Fig.~\ref{beta}, where the projection of the probability distribution $P(\kappa_j)$ in the final state (or at $t_\mathrm{max}$) is depicted. 
For a low initial mean diffusion coefficient of the system, e.g., for $\kappa = 10^{-5}$, we have observed that increasing slightly $\beta$,
%$\beta > 0$ leads to the appearance of the maximum of $P(\kappa_j)$ related with the cluster invasion at large, but smaller than $2 \kappa$, values of $\kappa_j$ (Fig.~\ref{beta}a).
%At the same time $P(\kappa_j)$ will maintain the maximum at small values of $\kappa_j$ induced by the faster density enhancement near the individuals with smaller $\kappa_j$ in the beginning of the time evolution.
%However, in general 
the distribution $P(\kappa_j)$ remains also now rather similar to the initial distribution, due to the reasons discussed already for $\beta = 0$.
However, for large values of $\beta$ a clear maximum of $P(\kappa_j)$  appears at large values of $\kappa_j$.
At intermediate values of $\kappa$ (e.g., $\kappa = 10^{-4}$), increasing the value of $\beta$ shifts $P(\kappa_j)$ to larger values (Fig.~\ref{beta}b), i.e., the individuals with larger diffusivities gain the competition advantage.
The transition from small to large values of the favorable diffusivities takes place smoothly.
Instead, for larger values of $\kappa$, e.g., $\kappa = 10^{-3}$, already a small increase of $\beta$ shifts the maximum of $P(\kappa_j)$ to the large values of $\kappa_j$ (Fig.~\ref{beta}c).
In all the cases, the increase of the competition success of the larger diffusivities is related to the fact that the larger fluctuations related to the increase of the death rates lead to the occasional disappearances of entire clusters or to their weakening and the organisms with larger diffusivities are more effective in occupying the empty space.
We also mention that increasing $\beta$ leads to smaller disappearance times, i.e., the final state of the system with identical organisms present is reached faster \cite{Heinsalu-2018}.
However, also the extinction probability of the total system increases \cite{Heinsalu-2012-PRE, Lande-1993, Legendre-1999}.

Increasing the competition intensity $\gamma = \alpha + \beta$ results in decreasing the carrying capacity of the system, which leads to the enhancement of the fluctuations in the population size.
Namely, the larger is $\gamma$ the smaller is the critical cluster size determined by the equilibrium between the reproduction and death rates [see Eq.~(\ref{NR*})] and the more probable is that a cluster disappears at some moment.
Thus, the effect is similar to the increase of $\beta$ keeping $\gamma$ constant: the organisms with large diffusivities will have the competition success.
This agrees well with the conclusion of Ref.~\cite{McPeek-1992} where a two-patch model was investigated, that a low-dispersing species dominates high carrying-capacity patches, whereas a high-dispersing species dominates low carrying-capacity patches.

\begin{figure*}[!t]
\includegraphics[width=15.0cm]{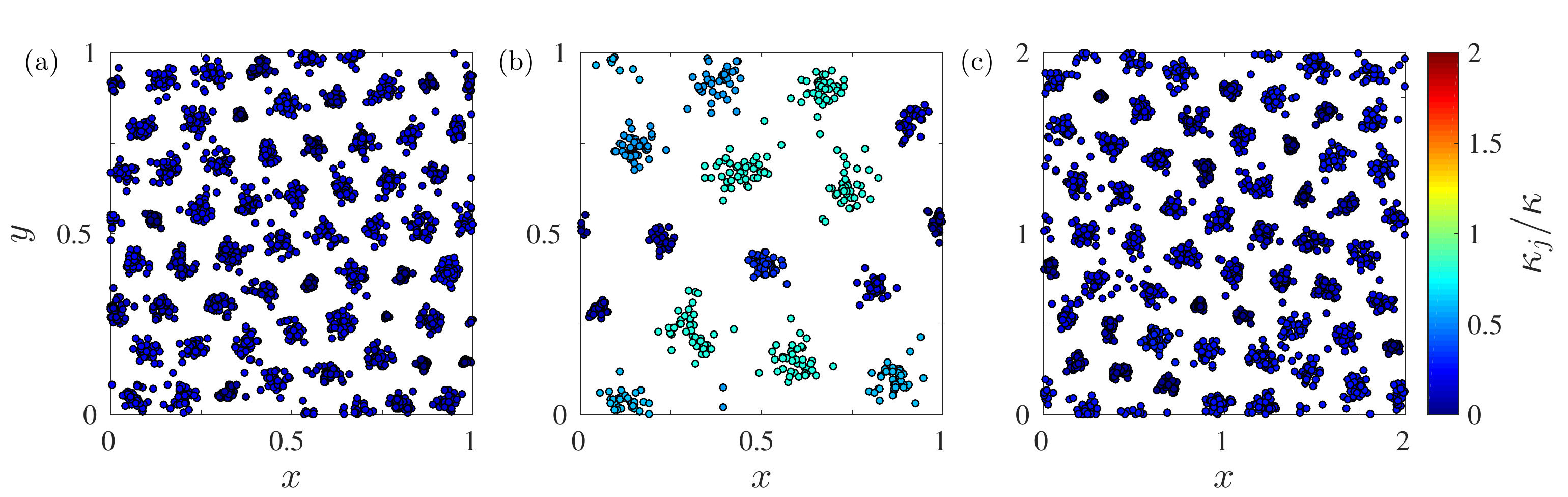}
\caption{Spatial configurations of Brownian bugs for different values of interaction radius in the two/dimensional system: (a) $R = 0.1$, $\kappa = 10^{-4}$, $L = 1$; (b) $R = 0.2$, $\kappa = 10^{-4}$, $L = 1$; (c) $R = 0.2$, $\kappa = 4 \times 10^{-4}$, $L = 2$. }
\label{XY-R}
\end{figure*}

We point out that though we have not assumed that a larger dispersal has a cost, this feature emerges in the systems where the clustering and pattern formation occurs naturally due to the neighborhood dependent reproduction rates (\ref{rates}).
Namely, for $\alpha > 0$, the larger is the diffusion coefficient of an individual the lower is on average its reproduction rate; the effect is the larger the stronger is the clustering (see the discussion in Ref.~\cite{EHG-2015}).
Furthermore, in the case of small temporal fluctuations, considering sexual reproduction and  Allee effect does not affect the results because the individuals gathered in clusters have many neighbors and the ones between the clusters even more. 
The Allee effect becomes important only for large temporal fluctuations when empty regions form due to the disappearance of clusters.

%%%%%%%%%%%%%%%%%%%%%%%%%%%%%%%%%%%%%%%%%%%%%

\subsection{Interaction radius} 

%%%%%%%%%%%%%%%%%%%%%%%%%%%%%%%%%%%%%%%%%%%%%

%
\begin{figure}[!b]
\includegraphics[width=7.5cm]{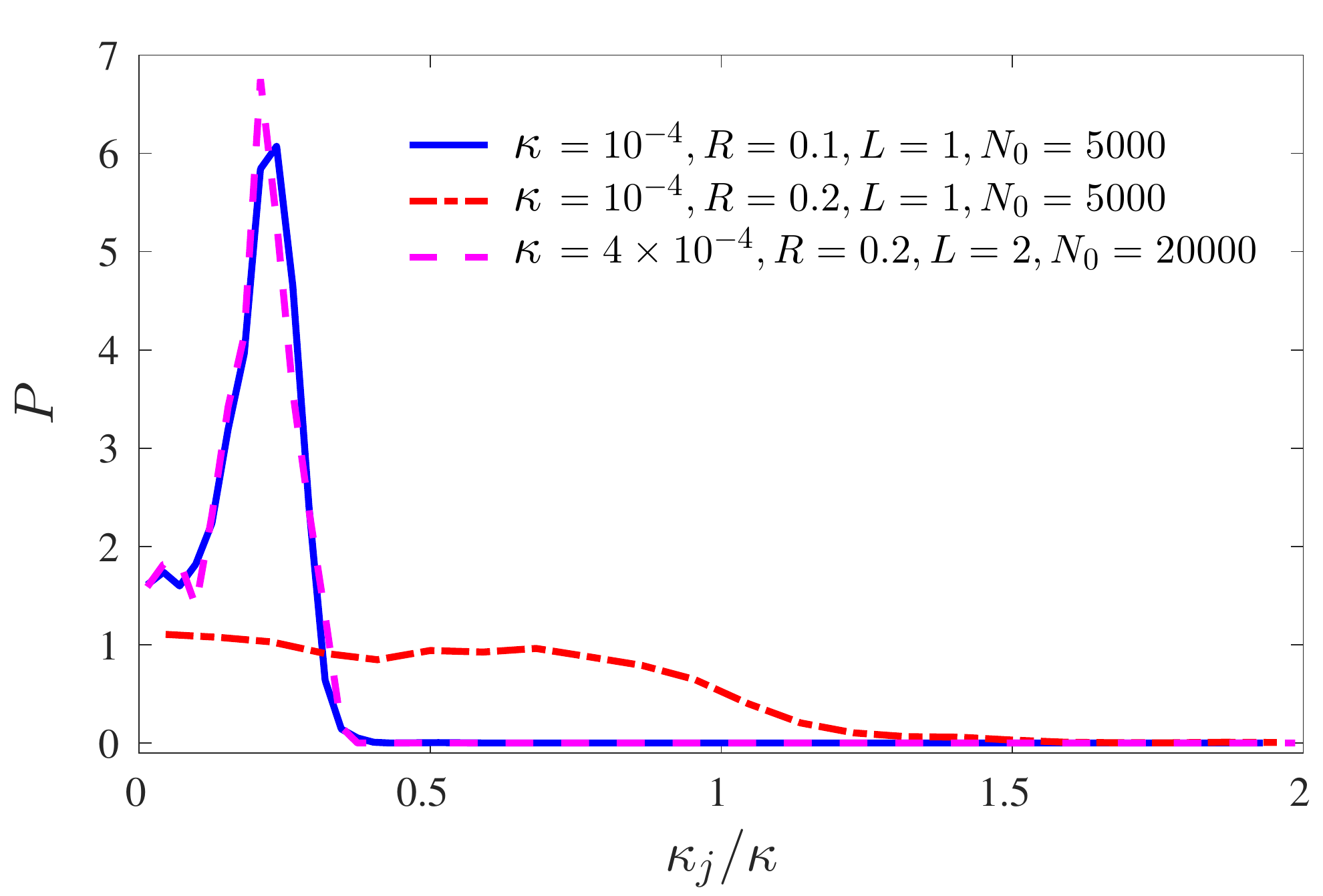}
\caption{The probability distribution of diffusivities $\kappa_j$ for different values of interaction radius $R$ and simulation box size $L$ in the two-dimensional system: (a) $R = 0.1$, $\kappa = 10^{-4}$, $L = 1$; (b) $R = 0.2$, $\kappa = 10^{-4}$, $L = 1$; (c) $R = 0.2$, $\kappa = 4 \times 10^{-4}$, $L = 2$. }
\label{P-R}
\end{figure}
\begin{figure}[!b]
\includegraphics[width=7.5cm]{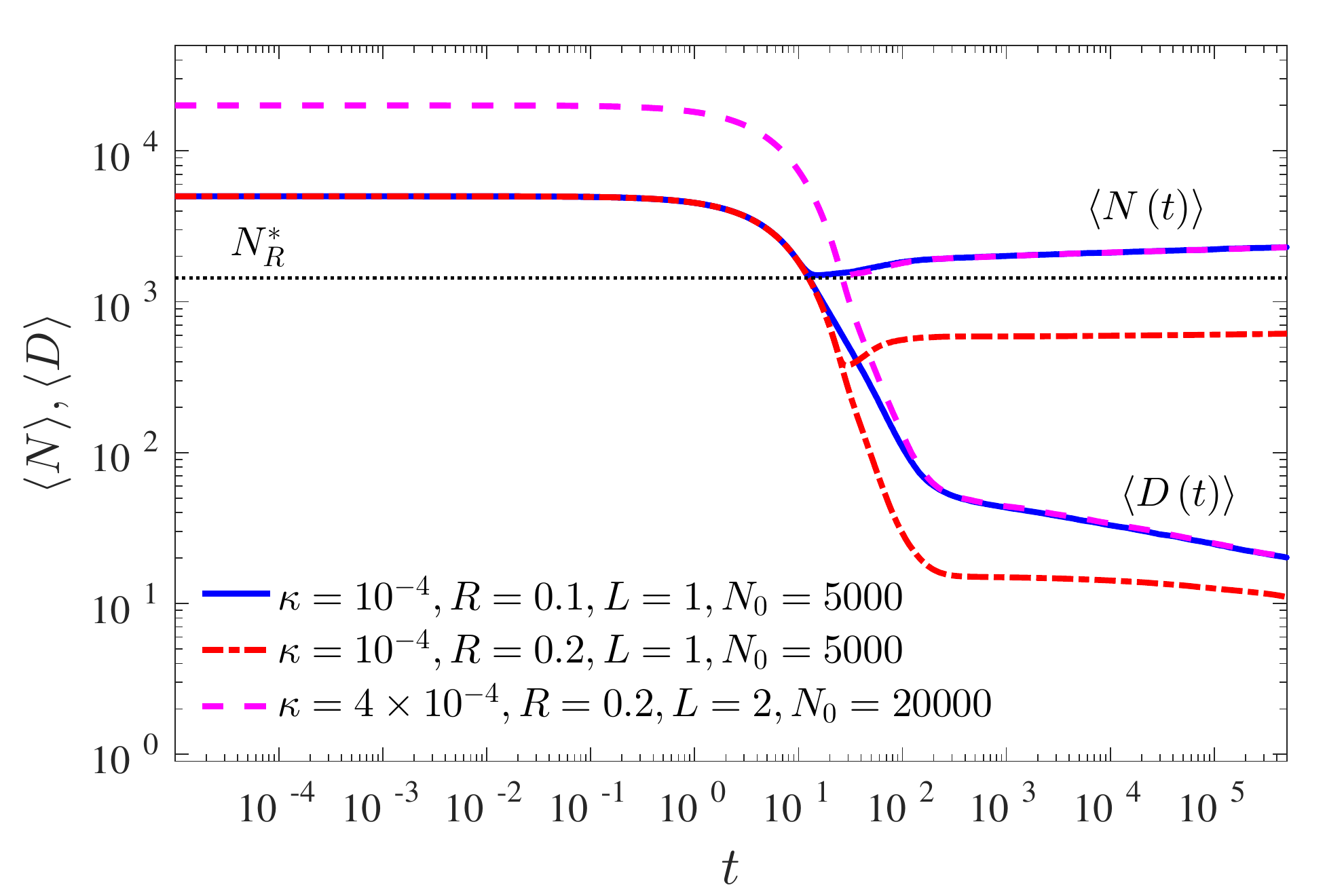}
\caption{The time evolution of average number of organisms and average diversity in the system for different values of interaction radius and simulation box size $L$ in the two-dimensional system: (a) $R = 0.1$, $\kappa = 10^{-4}$, $L = 1$; (b) $R = 0.2$, $\kappa = 10^{-4}$, $L = 1$; (c) $R = 0.2$, $\kappa = 4 \times 10^{-4}$, $L = 2$. }
\label{DN-R}
\end{figure}

The large-scale collective behavior of the system is first and foremost influenced by the competitive interaction \cite{Heinsalu-2010-EPL} and the periodicity of the spatial pattern formed is determined by the interaction radius. 
Increasing $R$ by a factor $r$ increases the pattern periodicity by the same factor, see Eq.~(\ref{delta}) and compare panels a and b in Fig.~\ref{XY-R} for the two-dimensional system.
The interaction radius enters also in the condition for pattern formation, Eq.~(\ref{pattern}), as well as in Eq.~(\ref{kappaopt}) determining together with the demographic parameters the optimal diffusion coefficient leading to the competition advantage.
Thus, one might think that $R$ is a crucial quantity and it is important to investigate its influence on results.

In fact, as can be seen from Fig.~\ref{P-R}, keeping the initial mean diffusivity of the system constant, the probability distribution $P(\kappa_j)$ at large times becomes much broader when increasing the interaction radius.
In this case both the optimal diffusion coefficient as well as the critical diffusion coefficient for which the periodic pattern is still observed are $r^2$ times larger compared to the case with smaller $R$ (see Eqs.~(\ref{pattern}) and (\ref{kappaopt})), leading also to the widening of the distribution.
The picture turns to be more similar to the one seen in Fig.~\ref{tevol-b0}a. %, i.e., the selectivity is disappearing.
Thus, increasing $R$ but keeping $\kappa$ the same, the effect is as if one decreases effectively the values of $\kappa_j$.
Namely, while the typical life time of a family that depends solely on the demographic parameters remains the same, the typical first-passage time increases by factor $r^2$, i.e. as if the organisms are slower.

Instead, increasing also $\kappa$ by factor $r^2$, i.e., using $rR$ and $r^2 \kappa$, the final probability distribution $P(\kappa_j)$ versus $\kappa_j / \kappa$ is the same as for $R$ and $\kappa$, see Fig.~\ref{P-R}.
In this case the typical first-passage time $t^{*}$ of a family is the same as in the case when using the values $R$ and $\kappa$ and qualitatively the system is the same (compare also panels a and c in Fig.~\ref{XY-R}).

Rescaling also the simulation box size, i.e. taking $L=2$, also the average number of organisms in the system remains the same, see Fig.~\ref{DN-R}.
In fact, in Fig.~\ref{DN-R} we have started with the initial number of organisms four times larger for the system with $R=0.2$ compared to the system with $R=0.1$, but reaching the state when the clusters are formed in the system, the time evolution of $\langle N \rangle$ and $\langle D \rangle$ is the same in both cases.

Therefore, with no loss of generality one can consider a single value of the interaction radius $R$, as we did in the present paper.

%%%%%%%%%%%%%%%%%%%%%%%%%%%%%%%%%%%%%%%%%%%%%

\section{Conclusion and discussion} 

%%%%%%%%%%%%%%%%%%%%%%%%%%%%%%%%%%%%%%%%%%%%%

In this paper we investigated the dynamics and the underlying mechanisms of natural selection in dispersal-structured populations.
Notably, in the model examined in the present paper, the spatial distribution of the organisms as well as the temporal fluctuations are generated by the individuals themselves through the density-dependence of the demographic processes.
In accordance with other works the model studied demonstrated that in general the clumping of the organisms favors the individuals diffusing slower and forming stronger clusters while the increase of temporal fluctuations enhances the competition success of the individuals diffusing faster.
However, we have shown that the extreme values of the diffusivities do not lead to the largest competition success.
Instead, in most cases there exists an optimal range of diffusion coefficients giving the competition advantage, determined by the interplay between various factors such as patch formation, temporal fluctuations, 
%initial conditions, 
and carrying capacity of the system.
Moreover, the time dynamics of the system and of the emergence of the probability distribution $P(\kappa_j)$ showing the maximum at intermediate diffusivities, is very interesting and related to the initial density enhancements due to the fluctuations as well as to the inter- and intra-cluster competition and self-organization of the organisms.
In the case of small temporal fluctuations, this observation agrees well with the conclusion made in Ref.~\cite{McPeek-1992} that some level of dispersal is favored by selection under almost all regimes of habitat variability, instead of the smallest diffusivity as predicted from the mean-field theory.
The results of this study are particularly relevant in the problems of the motion of micro-organisms such as bacteria when the ability of an organism to move is determined by various factors such as its size, shape, etc. \cite{Butenko-2012, Mitchell-1992}, but may give useful references also for the behavior of the systems consisting of macro-organisms \cite{Saastamoinen-2018a}.

%%%%%%%%%%%%%%%%%%%%%%%%%%%%%%%%%

\subsection*{Acknowledgments} \label{Sec-conclusion}

%%%%%%%%%%%%%%%%%%%%%%%%%%%%%%%%%

This work was supported by institutional research funding IUT (IUT39­-1) of the Estonian Ministry of Education and Research, by the Estonian Research Council grant PUT (PUT1356), and EU through the European Regional Development Fund (ERDF) Center of Excellence (CoE) program grant TK133.
E.H. is grateful to E. Hern\'andez-Garc\'ia for useful comments.

\subsection*{Authors contributions statement}
All the authors were involved in writing the code. D.N.M. performed the numerical simulations and data analysis. E.H. made the analytical 
calculations. All the authors were involved in the preparation of the manuscript. All the authors have read and approved the final manuscript.

%
%% Non-BibTeX users please use

%\newpage
%\clearpage
%\bibliography{references_dnavidad}

\begin{thebibliography}{37}





\bibitem{Okubo-Levin} 
A.~Okubo, S.~A. Levin, second ed., Springer-Verlag New York, (2001)


\bibitem{Lewis-book}
M.~A. Lewis, P.~K. Maini, S.~Petrovskii, Springer-Verlag, New York, (2012)


\bibitem{Pigolotti-2014-PRL}
S.~Pigolotti, R.~Benzi, Phys. Rev. Lett. \textbf{112}, 188102 (2014)


\bibitem{Kessler-2009}
D.~Kessler, L.~Sander, Phys. Rev. E \textbf{80}, 041907 (2009)


\bibitem{Waddell-2010}
J.~Waddell, L.~Sander, C.~Doering, Theor. Popul. Biol. \textbf{77}, 279 (2010)


\bibitem{Novak-2014}
S.~Novak, Ecol. Evol. \textbf{4}, 4589 (2014)


\bibitem{Johnson-1990}
M.~Johnson, M.~Gaines, Annu. Rev. Ecol. Evol. Syst. \textbf{21}, 449 (1990)


\bibitem{Lin-2015}
Y.~Lin, H.~Kim, C.~Doering, J. Math. Biol. \textbf{70}, 647 (2015)


\bibitem{Heinsalu-2013-PRL}
E.~Heinsalu, E.~Hern\'andez-Garc\'ia, C.~L\'opez, Phys. Rev. Lett.
  \textbf{110}, 258101 (2013)


\bibitem{Hastings-1983}
A.~Hastings, Theor. Popul. Biol. \textbf{24}, 244 (1983)


\bibitem{Holt-1985}
M.~A. McPeek, R.~D. Holt, Theor. Popul. Biol. \textbf{28}, 181, (1985)


\bibitem{Dockery-1998}
J.~Dockery, V.~Hutson, K.~Mischaikow, M.~Pernarowski, J. Math. Biol.
  \textbf{37}, 61 (1998)


\bibitem{Hutson-2003}
V.~Hutson, S.~Martinez, K.~Mischaikow, G.~Vickers, J. Math. Biol. \textbf{47},
  483 (2003)

  
\bibitem{Dieckmann-1999}
U.~Dieckmann, B.~O'Hara, W.~Weisser, Trends Ecol. Evol. \textbf{14}, 88 (1999)
  
  
\bibitem{Hutson-2001}
V.~Hutson, K.~Mischaikow, P.~Polacik, J. Math. Biol. \textbf{43}, 501 (2001)


\bibitem{Baskett-2007}
M.~Baskett, J.~Weitz, S.~Levin, Am. Nat. \textbf{1}, 59 (2007)
  
  
\bibitem{Stevens-2010}
V.M. Stevens, C.~Turlure, M.~Baguette, Biol. Rev. \textbf{85}, 625 (2010)


\bibitem{EHG-2004}
E.~Hern\'andez-Garc\'ia, C.~L\'opez, Phys. Rev. E \textbf{70}, 016216 (2004)


\bibitem{Heinsalu-2018}
E.~Heinsalu, D.~{Navidad Maeso}, M.~Patriarca, Eur. Phys. J. B \textbf{91},
  255 (2018)


\bibitem{Saastamoinen-2018a}
M.~Saastamoinen, et~al., Biol. Rev.  \textbf{93},
  574 (2018)

  
\bibitem{Heinsalu-2012-PRE}
E.~Heinsalu, E.~Hern\'andez-Garc\'ia, C.~L\'opez, Phys. Rev. E \textbf{85},
  041105 (2012)
  
  
\bibitem{Heinsalu-2010-EPL}
E.~Heinsalu, E.~Hern\'andez-Garc\'ia, C.~L\'opez, Europhys. Lett. \textbf{92},
  40011 (2010)
  
  
\bibitem{CL-2004}
C.~L\'opez, E.~Hern\'andez-Garc\'ia, Physica D \textbf{199}, 223 (2004)

  
  
\bibitem{EHG-2015}
E.~Hern\'andez-Garc\'ia, E.~Heinsalu, C.~L\'opez, Ecol. Complex. \textbf{21},
  166 (2015)


\bibitem{EHG-2005}
E.~Hern\'andez-Garc\'ia, C.~L\'opez, J. Phys.: Condens. Matter \textbf{17},
  S4263 (2005)
  

 
\bibitem{Cross-1993}
M.~C. Cross, P.~C. Hohenberg, Rev. Mod. Phys. \textbf{65} 851 (1993)


\bibitem{Hamilton-1977}
W.~D. Hamilton, R.~M. May, Nature \textbf{269} 578 (1977)

  
  
\bibitem{Comins-1980}
H.~N. Comins, W.~D. Hamilton, R.~M. May, J. Theor. Biol. \textbf{82} 205 (1980)

  
  
\bibitem{Frank-1986}
S.~A. Frank, J. Theor. Biol. \textbf{122} 303 (1986)

  
\bibitem{Gandon-1999}
S.~Gandon, Y.~Michalakis, J. Theor. Biol. \textbf{199} 275 (1999)

  
\bibitem{Heino-2001}
M.~Heino, I.~Hanski, American Naturalist \textbf{157}, 495 (2001)


\bibitem{Dyken-2010}
J.~D.~V. Dyken, Evolution, American Naturalist \textbf{64}, 2840 (2010)


\bibitem{Lande-1993}
R.~Lande, Evolution, Am. Nat. \textbf{142}, 911 (1993)


\bibitem{Legendre-1999}
S.~Legendre, J.~Clobert, A.~Moller, G.~Sorci, Am. Nat. \textbf{153}, 449 (1999)


\bibitem{McPeek-1992}
M.~A. McPeek, R.~D. Holt, Am. Nat. \textbf{140}, 1010 (1992)


\bibitem{Butenko-2012}
A.~V. Butenko, E.~Mogilko, L.~Amitai, B.~Pokroy, E.~Sloutskin, Langmuir \textbf{28}, 12941 (2012)



\bibitem{Mitchell-1992}
J.~G. Mitchell, Microb. Ecol. \textbf{22}, 227 (1991)


\end{thebibliography}
%\bibliographystyle{plain}

\end{document}